\documentclass[conference]{IEEEtran}
\IEEEoverridecommandlockouts

\usepackage{cite}
\usepackage{amsmath,amssymb,amsfonts}
\usepackage{algorithmic}
\usepackage{graphicx}
\usepackage{textcomp}
\usepackage{xcolor}
\usepackage{url}
\usepackage[colorlinks=true, allcolors=blue]{hyperref}
\def\BibTeX{{\rm B\kern-.05em{\sc i\kern-.025em b}\kern-.08em
    T\kern-.1667em\lower.7ex\hbox{E}\kern-.125emX}}
\begin{document}

\title{A Systems Design Framework for Hybrid Quantum–Classical Methods in Density Functional Theory}

\author{\IEEEauthorblockN{
Namrata Manglani\IEEEauthorrefmark{1}\IEEEauthorrefmark{2},
Samrit Kumar Maity\IEEEauthorrefmark{1},
Shashank Sharma\IEEEauthorrefmark{1},
Soham Phulare\IEEEauthorrefmark{1},
Sanjay Wandhekar\IEEEauthorrefmark{1}
}

\IEEEauthorblockA{\IEEEauthorrefmark{1}
C-DAC, Pune, India
}

\IEEEauthorblockA{\IEEEauthorrefmark{2}
Shah and Anchor Kutchhi Engineering College, Mumbai, India
}

\IEEEauthorblockA{
Email: namrata.manglani@sakec.ac.in, samritm@cdac.in, shashank.sharma@cdac.in, sohamphulare05@gmail.com sanjayw@cdac.in
}

}

\maketitle

\begin{abstract}
Density Functional Theory (DFT) is the workhorse of atomistic simulation, yet large-scale electronic-structure calculations remain limited by computational cost and the accuracy of approximate exchange correlation functionals. Although quantum computing approaches offer promising solutions, including variational techniques, embedding strategies, and quantum linear solvers, the discussion remains largely scattered. Without shared terms or structure, the evaluation of progress in hybrid quantum-classical frameworks for DFT (HQ-DFT) becomes challenging. We present the first systematic design framework for HQ-DFT, organizing methodologies according to (i) the level of quantum intervention within the DFT workflow and (ii) the highest stage of validation (theory, simulation, or hardware demonstration). Existing HQ-DFT methodologies can be represented as a point in this design space, providing a common basis for comparison and analysis. 
Mapping representative studies onto this design space reveals that Hamiltonian- and partitioning-based approaches dominate current implementations, whereas solver-level acceleration remains largely confined to theoretical studies because of its reliance on fault-tolerant quantum computing. The resulting framework provides reusable systems-level design methodology for comparing, designing, and implementing HQ-DFT workflows, while serving as a practical decision-support tool for future hybrid quantum-classical electronic-structure workflows.
\end{abstract}

\begin{IEEEkeywords}
Density Functional Theory, Quantum Computing, Self-Consistent Field, Quantum Algorithms, Electronic Structure, Hybrid Quantum-Classical Methods, Quantum Chemistry Software.
\end{IEEEkeywords}

\section{Introduction}

Density functional theory (DFT) is known to balance accuracy and computational cost across numerous applications \cite{kohn1999nobel,jones2015density}. Yet this balance breaks down under certain conditions already recognized in the field \cite{becke1993density,davidson1975iterative, kotliar2006electronic}. Consequently, the development of improved electronic-structure methodologies remains an active area of research.

Quantum algorithms provide a native representation of many-electron wavefunctions which are difficult to capture classically and hence show promise for resolution of some of DFT's significant limitations\cite{peruzzo2014,reiher2017elucidating,babbush2023encoding,ko2023,sun2016quantum}. However, continued reliance on DFT for practical electronic-structure calculations have motivated hybrid workflows, where DFT and quantum algorithms are used together. 

Existing research at the intersection of quantum computing and electronic-structure theory follows two distinct directions: quantum-native methods that replace classical electronic-structure calculations, and DFT-centric hybrid methods that augment selected stages of the Kohn–Sham workflow. Despite their fundamentally different objectives and computational workflows, these approaches are often discussed together, making systematic comparison and evaluation difficult.

To distinguish this emerging class of methods, we use the term Hybrid Quantum–Classical methods for Density Functional Theory (HQ-DFT) to denote DFT-centric methodologies in which quantum computation augments, rather than replaces, the Kohn–Sham workflow. Building on this definition, we propose a two-dimensional framework in (Section~\ref{sec:frame}). The framework is applied via systematic mapping of representative literature (Section~\ref{sec:validate}), from which we derive practical design guidelines (Section~\ref{sec:design}) and implementation recommendations for HQ-DFT software ecosystems (Section~\ref{sec:software}).

\section{Motivation for HQ-DFT}
\label{sec:dft}

The practical success of Density Functional Theory (DFT) is tied to the use of approximations that enable computational efficiency but introduce systematic limitations in electronically complex systems. These failures can be categorized into intrinsic functional errors and computational scaling bottlenecks.

\textit{Intrinsic Limitations:} The exact exchange-correlation (XC) functional is unknown. All practical approximations, such as LDA, GGA, and hybrid functionals, suffer from self-interaction and delocalization errors, which produce incorrect charge localization and underestimated band gaps \cite{Perdew1981_SIC,Cohen2008_SIE}. Furthermore, $f$-electron compounds and transition-metal oxides, where Coulomb correlations are strong, are poorly handled by standard functionals \cite{Cohen2012_StrongCorrelation}. While Time-Dependent DFT (TD-DFT) extends the theory to excited states, it fails to accurately capture charge-transfer excitations and long-time dynamics \cite{RungeGross1984_TDDFT,Maitra2016_TDDFTLimitations}.

\textit{Computational Bottlenecks:} The cost of solving the Kohn-Sham (KS) equations is dominated by the repeated diagonalization of the Hamiltonian matrix, which scales as $O(N^3)$ relative to the system size $N$ \cite{Goedecker1999_Scaling}. Use of hybrid or double-hybrid functionals, improve accuracy but at a cost of $O(N^4)$ to $O(N^5)$, rendering them prohibitive for large-scale correlated materials.

These two classes of limitations motivate different forms of hybrid quantum assistance. Some approaches seek to improve the physical description of electronic interactions, others accelerate computational kernels, while others restrict quantum computation to localized regions of interest. As these strategies intervene at different stages of iterative Kohn-Sham DFT workflow, a systematic framework is required to organize, compare, and guide the design of HQ-DFT methodologies.

\section{A Structured Framework for HQ-DFT}
\label{sec:frame}
\begin{figure}[htbp]
\centering
\includegraphics[width=0.8\columnwidth]{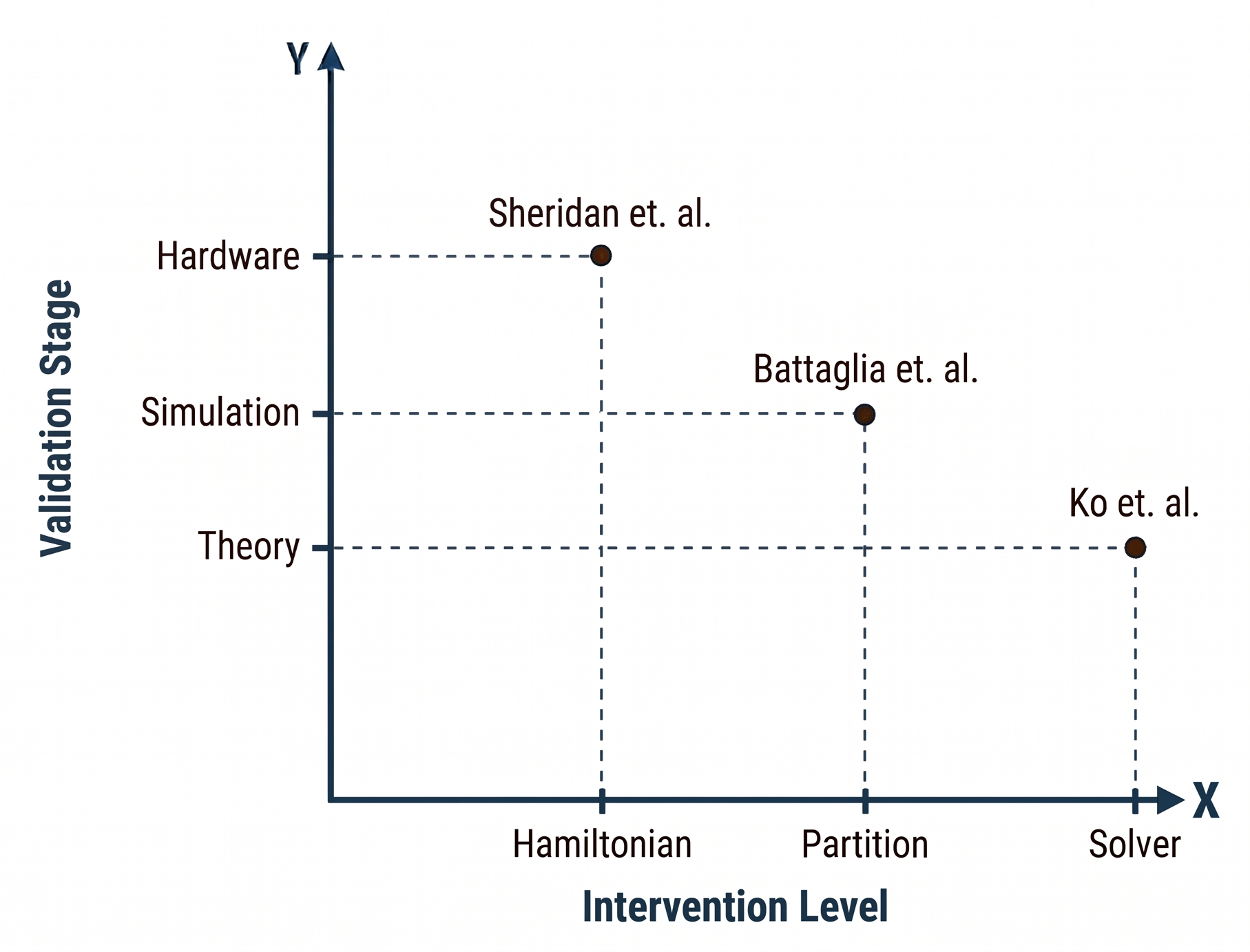}
\caption{\textbf{Design space for HQ-DFT each point is a representative study.} }
\label{fig:hq_dft_cone}
\end{figure}

We propose a systems design methodology for Hybrid Quantum–Classical Density Functional Theory (HQ-DFT) that abstracts hybrid electronic-structure workflows using two orthogonal design dimensions, as illustrated in Figure~\ref{fig:hq_dft_cone}. The framework provides a reusable abstraction for selecting quantum intervention strategies, implementation pathways, and validation objectives during the design and development of hybrid quantum–classical electronic-structure systems. The intervention dimension identifies where quantum resources augment the Kohn–Sham workflow, while the validation dimension captures the maturity of an approach from theoretical development through simulation to hardware demonstration. Together, these dimensions provide a structured basis for designing, comparing, and implementing HQ-DFT workflows, independent of the underlying quantum algorithm, software stack, or hardware platform.

\subsection{Intervention Level}
\begin{figure}[htbp]
\centering
\includegraphics[width=0.95\columnwidth]{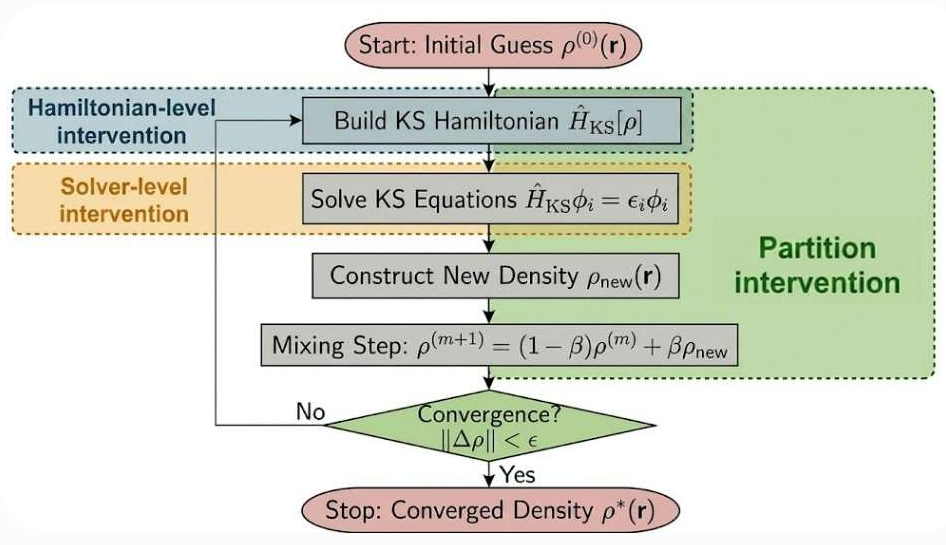}
\caption{\textbf{Intervention levels in the KS-DFT workflow.}}
\label{fig:scf}
\end{figure}

The intervention levels are derived from the structure of DFT self- consistent workflow rather than the surveyed literature. The stage at which quantum resources are integrated into the workflow determines the intervention level. As shown in Figure \ref{fig:scf} if quantum resources improve the Hamiltonian construction, it is Hamiltonian-level intervention; if they are used only in the eigensolver, it is a solver-level intervention; if they enter multiple stages of the SCF workflow by working on a part of system, it is a partition intervention.

\paragraph{Hamiltonian-level Intervention.}
Hamiltonian-level interventions are characterized by the modification of the core mathematical terms within the electronic structure problem prior to the self-consistent field (SCF) loop. These methods utilize quantum-derived correlation data to directly refine the classical Kohn-Sham (KS) Hamiltonian. 

A primary example is the work of Hatcher et al. \cite{Hatcher2019}, where a quantum computer generates high-fidelity exchange-correlation (XC) potentials to ensure formal exactness within the classical DFT cycle. Similarly, Sheridan et al. \cite{Sheridan2024} propose Quantum Enhanced DFT (QE-DFT), which leverages quantum data to optimize the underlying parameters of XC functionals for improved accuracy in noisy regimes. This category  determines the exact KS potentials through epigraphical optimization \cite{baker2020}. Any strategy that employs quantum resources to overcome intrinsic limitations of DFT or parameterize the classical potential belongs to this intervention level. Performance wise this intervention ensures improvement in the model description, such that the classical mean-field context is grounded in high-level quantum correlation.

\paragraph{Solver level.}
Solver-level interventions replace or augment the numerical electronic structure solver within the self-consistent field (SCF) procedure while preserving the underlying electronic Hamiltonian. Unlike Hamiltonian-level approaches, which improve the physical description of the system, solver interventions target the computational bottlenecks associated with repeated eigenvalue calculations, density-matrix construction, and iterative SCF convergence. These operations dominate the computational cost for large molecules, periodic materials, and condensed-matter systems.

Near-term implementations predominantly employ Variational Quantum Eigensolvers (VQE), which utilize shallow parameterized quantum circuits combined with classical optimization to approximately solve the Kohn-Sham eigenvalue problem while remaining compatible with noisy quantum hardware \cite{peruzzo2014,mcclean2017}. Beyond NISQ devices, fault-tolerant algorithms such as Quantum Phase Estimation (QPE) and Quantum Singular Value Transformation (QSVT) offer asymptotic improvements for eigenvalue estimation and density-matrix purification, potentially reducing the computational complexity of large-scale DFT calculations \cite{Aspuru2005,ko2023}. Consequently, solver-level interventions are particularly suited to applications where the electronic structure model is sufficiently accurate but the numerical solution of the SCF equations becomes the dominant computational bottleneck.

\paragraph{Partitioning Intervention.}
Partitioning interventions decompose the electronic system into a localized active region requiring high-accuracy quantum treatment and a surrounding environment that can be efficiently described using classical DFT. Unlike Hamiltonian-level interventions, which improve the physical model, or solver-level interventions, which accelerate numerical solution procedures, partitioning interventions reduce the computational resources required by restricting quantum computation to the chemically significant portion of the system.

The theoretical foundation for this approach is provided by quantum embedding methods, which couple an accurate quantum description of the active region with a classical mean-field representation of the environment \cite{sun2016quantum}. By limiting the size of the quantum subsystem, partitioning substantially reduces qubit requirements, circuit depth, and measurement overhead while preserving the accuracy of the overall calculation.

Representative HQ-DFT implementations include the project-based HF/DFT embedding framework developed by Rossmannek \textit{et al.} \cite{rossmannek2020,rossmannek2023}, the active-space embedding framework proposed by Battaglia \textit{et al.} \cite{Battaglia2024}, the range-separated DFT embedding approach of Poirier \textit{et al.} \cite{Poirier2024}, and the DFT-embedded post-processed QSCI methodology introduced by Do \textit{et al.} \cite{Do2026}. These methods enable simulations of large molecular and condensed-matter systems, including catalytic surfaces, crystalline defects, and metal--organic frameworks, where strong electron correlation remains spatially localized. Consequently, partitioning interventions provide the most scalable strategy for extending HQ-DFT to realistic large-scale applications while maintaining the efficiency of the surrounding classical DFT workflow.

\subsection{Validation Stage}

The validation stage is the highest level at which an HQ-DFT approach is demonstrated in studies,which indicates its technological maturity. Mostly studies progress from theoretical development to simulation and eventually hardware implementation, each work is classified according to its highest reported validation stage.

\paragraph{Theory}
Theory-level studies propose novel HQ-DFT algorithms or analytical models without any experimental validation. They establish the conceptual foundations of approaches and provide theoretical expectations of results.

\paragraph{Simulation}
Simulation-level studies are the ones that validate proposed methods using classical quantum simulators. These investigations check correctness of algorithms , numerical accuracy, and scalability under noiseless or noise-modeled simulation environment, before the approach is deployed physical quantum hardware.

\paragraph{Hardware}
Hardware-level studies show results of HQ-DFT methods executed on real quantum devices, such as IBM Quantum, Rigetti, IonQ, Quantinuum, or other available platforms. They evaluate the practical feasibility of the proposed approaches under real hardware constraints, including noise, limited qubit counts, coherence times and device connectivity.

For classification purpose, each study has been assigned to its \emph{highest} reported validation stage (Hardware $>$ Simulation $>$ Theory), ensuring a  consistent and unique mapping inside the proposed framework.

\section{Framework-Based Analysis of HQ-DFT Literature}
\label{sec:validate}

The following analysis of the proposed HQ-DFT framework is demonstrated by its ability to categorize the diverse landscape of the existing literature into the proposed design space. Table~\ref{tab:mapping} maps representative methodologies across the two dimensions of the framework. 

\begin{table*}[!t]
\renewcommand{\arraystretch}{1.3}
\caption{Representative HQ-DFT studies mapped onto the proposed design space.}
\label{tab:mapping}
\centering
\begin{tabular}{p{10cm} c c}
\hline
\bfseries Description and Justification & \bfseries Intervention Level & \bfseries Validation Stage\\
\hline
Hatcher et al. (2019) incorporate quantum-computed correlation energy into the Kohn--Sham Hamiltonian during the SCF cycle \cite{Hatcher2019}. & Hamiltonian & Theory \\
Rossmannek et al. (2021) partition electronic systems into quantum active spaces embedded within a classical mean-field environment \cite{rossmannek2020}. & Partitioning & Simulation \\
Baker and Poulin (2021) develop a machine-learned exact exchange--correlation functional using quantum-generated data \cite{baker2020}. & Hamiltonian & Theory \\
Rossmannek et al. (2023) implement a quantum embedding framework for strongly correlated electronic systems on quantum hardware \cite{rossmannek2023}. & Partitioning & Hardware \\
Senjean et al. (2023) formulate a theoretical framework for performing density functional theory on quantum computers \cite{Senjean2023}. & Hamiltonian & Theory \\
Ko et al. (2023) propose a quantum algorithm based on QSVT for self-consistent density functional theory \cite{ko2023}. & Solver & Theory \\
Sheridan et al. (2024) optimize exchange--correlation functional parameters using quantum-generated data to improve DFT accuracy \cite{Sheridan2024}. & Hamiltonian & Hardware \\
Poirier et al. (2024) integrate range-separated density functional theory with multiresolution analysis and quantum computing \cite{Poirier2024}. & Partitioning & Simulation \\
Traore et al. (2024) employ quantum basis-set correction to improve chemical accuracy in density functional calculations \cite{traore2024}. & Hamiltonian & Simulation \\
Battaglia et al. (2024) present an active-space embedding framework for hybrid quantum--classical electronic structure calculations \cite{Battaglia2024}. & Partitioning & Simulation \\
Sokolov et al. (2026) introduce quantum-enhanced neural exchange--correlation functionals for improved electronic structure prediction \cite{sokolov2026}. & Hamiltonian & Simulation \\
Do et al. (2026) combine DFT embedding with post-processed quantum selected configuration interaction for large-scale electronic-structure calculations \cite{Do2026}. & Partitioning & Hardware \\
\hline
\end{tabular}
\end{table*}

 The classification in Table~\ref{tab:mapping} reveals several important trends in the evolution of HQ-DFT. Hamiltonian-level interventions dominate the current literature, reflecting the emphasis on improving the physical accuracy of approximate exchange--correlation functionals. More recent studies increasingly adopt partitioning strategies to improve scalability by confining quantum computation to localized strongly correlated regions. In contrast, solver-level interventions remain limited and are primarily theoretical, indicating that quantum acceleration of self-consistent field (SCF) computations remains an important open research direction.

The proposed framework also reveals a correspondence between intervention level and the underlying computational bottleneck. Hamiltonian interventions address model limitations, solver interventions target computational complexity, and partitioning interventions overcome quantum resource constraints through active-region decomposition. Consequently, the framework serves as a practical design aid, enabling researchers to select intervention strategies according to the dominant bottleneck while highlighting underexplored directions for future HQ-DFT research.

\subsection{Framework-derived Design Guidelines}
\label{sec:design}
Beyond categorizing existing HQ-DFT studies, the proposed framework reveals that the choice of intervention level is determined by the dominant computational bottleneck rather than the quantum algorithm itself. Table~\ref{tab:designguidelines} summarizes this relationship and provides practical guidance for selecting an appropriate hybrid quantum--classical strategy.

\begin{table*}[!t]
\renewcommand{\arraystretch}{1.3}
\caption{Guidelines for selecting the appropriate intervention in HQ-DFT}
\label{tab:designguidelines}
\centering
\begin{tabular}{p{6cm}c p{6cm}}
\hline
\bfseries Problem Scenario & \bfseries Preferred Intervention & \bfseries Representative Applications (Example Studies) \\
\hline
Approximate exchange--correlation functionals fail to accurately describe strong electron correlation or electron localization. & Hamiltonian & H$_2$ dissociation, N$_2$ bond breaking, H$_8$ chain, Hubbard models \cite{Hatcher2019,Senjean2023,Sheridan2024}. \\
Expensive eigensolver and density-matrix construction & Solver & H$_2$O sheet, BaTiO$_3$ supercells \cite{ko2023}\\
Localized correlation within a larger environment & Partitioning & Crystalline defects, catalytic active sites, and metaloxide frameworks (e.g., MgO oxygen vacancy, HKUST-1, CO adsorption on HEA nanoparticles) \cite{rossmannek2023,Battaglia2024,Do2026}. \\
\hline
\end{tabular}
\end{table*}

The framework demonstrates that intervention strategies naturally align with three distinct classes of computational bottlenecks encountered in DFT. Hamiltonian interventions address deficiencies in the underlying electronic structure model, solver interventions accelerate computationally intensive SCF kernels without altering the physical model, and partitioning interventions improve scalability by restricting quantum computation to localized strongly correlated regions. Consequently, the proposed framework functions as a decision-support methodology that guides the selection of intervention strategies according to the characteristics of the target application while highlighting underexplored opportunities, particularly in quantum-accelerated SCF solvers.

\textbf{Key Insight:} The proposed intervention levels correspond to three fundamentally different classes of computational bottlenecks in DFT model limitations (Hamiltonian), algorithmic limitations (Solver), and resource limitations (Partitioning). This correspondence provides a principled basis for selecting hybrid quantum--classical strategies according to the characteristics of the target application rather than the choice of quantum algorithm.

\section{Software Readiness and Implementation Realities}
\label{sec:software}

The proposed framework identifies \emph{where} quantum computation is introduced into the DFT workflow. Practical implementation, however, also depends on the material class under investigation, since molecular and periodic systems employ fundamentally different electronic-structure representations and software ecosystems. Consequently, software selection is governed jointly by the target application and the chosen intervention level. Table~\ref{tab:software_readiness} summarizes the implementation stacks most commonly adopted by the representative HQ-DFT studies surveyed in this work.

\begin{table*}[!t]
\renewcommand{\arraystretch}{1.3}
\caption{Representative implementation stacks observed for different HQ-DFT application scenarios.}
\label{tab:software_readiness}
\centering
\begin{tabular}{l c l}
\hline
\bfseries Material Class & \bfseries Typical HQ-DFT Strategy & \bfseries Representative Implementation Stack \\
\hline
Molecular systems & Hamiltonian, Partitioning & PySCF \cite{pyscf_2018} $\rightarrow$ Qiskit Nature \cite{qiskit_2017} $\rightarrow$ Qiskit \cite{qiskit_2017} \\
Large periodic materials & Solver & M-SPARC \cite{msparc_2021} $\rightarrow$ Sparse Oracle + QSVT \cite{ko2023}$^{\mathrm{a}}$ \\
Condensed matter, defects, MOFs & Partitioning & CP2K \cite{cp2k_2020} $\rightarrow$ Qiskit Nature \cite{qiskit_2017} $\rightarrow$ Qiskit \cite{qiskit_2017} \\
\hline
\multicolumn{3}{l}{$^{\mathrm{a}}$ Representing a theoretical framework lacking a native SDK integration wrapper.}
\end{tabular}
\end{table*}

The implementation stacks in Table~\ref{tab:software_readiness} are derived from the surveyed HQ-DFT literature and should therefore be interpreted as representative rather than prescriptive. PySCF has emerged as the dominant molecular electronic-structure platform owing to its modular Python interface, direct access to molecular integrals, and seamless integration with quantum computing frameworks such as Qiskit Nature, making it the preferred environment for Hamiltonian development and quantum embedding workflows \cite{pyscf_2018,pyscf_2020}. For large periodic materials, the only reported solver-level HQ-DFT implementation employs M-SPARC, whose scalable real-space DFT infrastructure enables direct replacement of the self-consistent field (SCF) solver while preserving the remainder of the electronic-structure workflow \cite{ko2023}. Similarly, CP2K has become the preferred platform for embedding-based studies involving condensed-phase systems, crystalline defects, and metal--organic frameworks because of its efficient Gaussian--Plane Wave (GPW) formulation and mature subsystem capabilities \cite{cp2k_2020}.

Several established electronic-structure packages are intentionally absent from the recommended implementation stacks. Psi4 provides high-quality molecular integral generation and an excellent Python interface, making it valuable for algorithm prototyping and benchmarking \cite{psi4_2017,psi4_2020}; however, the surveyed HQ-DFT literature overwhelmingly adopts PySCF for molecular Hamiltonian and embedding development because of its stronger interoperability with quantum software frameworks. Likewise, although Quantum ESPRESSO is one of the most widely used plane-wave DFT packages for periodic materials, none of the representative HQ-DFT studies surveyed in this work employs it as the primary implementation platform. Accordingly, it is not included in the recommended implementation stacks, emphasizing that the recommendations presented here are evidence-driven and reflect current practice rather than the broader capabilities of individual electronic-structure packages.

\paragraph{Implementation Insights.}
 The framework reveals that practical HQ-DFT design requires balancing three competing factors. Hamiltonian interventions maximize physical accuracy, but often increase quantum circuit complexity. Solver interventions offer the greatest potential computational acceleration but currently depend on fault-tolerant quantum algorithms, limiting near-term deployment. Partitioning interventions provide a practical route for large-scale systems by reducing quantum resource requirements, although the choice of active region directly influences accuracy. Consequently, successful HQ-DFT implementations require jointly selecting the intervention strategy, software ecosystem, and quantum hardware according to the characteristics of the target application.

 \paragraph{Systems-level implementation considerations.}
The practical realization of HQ-DFT workflows is influenced by algorithmic choices and systems-level characteristics of hybrid computing platforms. In addition to the computational considerations discussed above, practical implementations must account for:

\begin{itemize}
    \item Memory and data movement: Hybrid HQ-DFT workflows require repeated exchange of information between classical and quantum components. The absence of a shared memory makes explicit movement and data synchronization necessary \cite{Battaglia2024}.

    \item Classical-quantum communication: Frequent interaction introduces communication latency that can dominate execution time, especially for partitioning-based approaches that involve repeated invocations of quantum subroutines \cite{Raj2026}.

    \item Runtime orchestration: Efficient execution requires coordinated scheduling classical computations with quantum tasks, making workflow orchestration and resource management an important systems consideration for scalable HQ-DFT implementations \cite{Luckow2023}.
   
\end{itemize}

These considerations complement the proposed framework by emphasizing that the choice of quantum intervention is influenced not only by the electronic-structure problem, but also by the characteristics of the underlying hybrid computing platform.

\section{Conclusion}
\label{sec:conclusion}

Hybrid Quantum-Classical methods for Density Functional Theory (HQ-DFT) are evolving through diverse strategies that integrate quantum computation into different stages of the electronic-structure workflow. However, the absence of a common design methodology has made it difficult to systematically compare these approaches and identify practical pathways for developing hybrid quantum-classical systems.

This work presents a two-dimensional HQ-DFT framework that captures both the point of quantum intervention within the DFT workflow and the validation maturity of an approach, from theoretical development to hardware demonstration. Applying the framework to representative studies reveals distinct design trends: Hamiltonian-level methods primarily improve electronic-structure accuracy, partitioning has emerged as the dominant strategy for scalable near-term implementations, and solver-level approaches remain an important long-term direction because of their reliance on fault-tolerant quantum computing. By relating computational bottlenecks to intervention strategies and representative software ecosystems, the framework provides practical guidance for selecting and implementing HQ-DFT workflows across molecular and periodic electronic-structure applications.

While the framework is derived from representative HQ-DFT studies, its practical utility is expected to evolve alongside advances in quantum hardware, software ecosystems, and hybrid execution models. The proposed methodology provides a reusable systems-level abstraction for designing, comparing, and implementing HQ-DFT workflows, offering a foundation for interoperable software stacks and the systematic development of next-generation hybrid quantum-classical electronic-structure applications.

\section*{Acknowledgment}

The author(s) gratefully acknowledge the support of the AICTE Industry Fellowship Scheme for funding and facilitating this research. The guidance and resources provided under this program have been invaluable in carrying out this work. The author(s) also thank the Centre for Development of Advanced Computing (C-DAC) and the National Supercomputing Mission (NSM) for providing technical support, that helped in completion this research.

\bibliographystyle{IEEEtran}
\bibliography{NG-QDFT}
\end{document}